\documentclass[runningheads]{llncs}

\usepackage[utf8]{inputenc} 
\usepackage[T1]{fontenc}    
\usepackage{hyperref}       
\usepackage{url}            
\usepackage{booktabs}       
\usepackage{amsfonts}       
\usepackage{nicefrac}       
\usepackage{microtype}      
\usepackage{xcolor}         
\usepackage{algorithm}
\usepackage{algpseudocode}
\usepackage{siunitx}
\usepackage{tabularx}

\usepackage{booktabs}
\usepackage{microtype}
\usepackage{graphicx}
\usepackage{subfigure}
\usepackage{booktabs} 
\usepackage{multirow}

\usepackage{hyperref}
\usepackage{caption}
\usepackage{xcolor}
\usepackage{color}

\usepackage{amsmath,amssymb,amscd,url,enumerate, mathtools}
\usepackage[utf8]{inputenc}
\usepackage{wrapfig}
\usepackage{pdflscape}
\usepackage{tikz}
\usetikzlibrary{decorations.markings,arrows}
\usetikzlibrary{positioning,arrows,fit}

\usepackage{tikz-cd}
\usepackage{amsfonts}
\usepackage{amsmath}
\usepackage{graphicx}
\usepackage{mathrsfs}
\usepackage{dsfont}
\usepackage{enumitem}
\usepackage{array}
\usepackage{verbatim}
\usepackage{ulem}
\newcolumntype{?}{!{\vrule width 1pt}}
\usepackage[capitalize,noabbrev]{cleveref}
\usepackage[font=footnotesize,labelfont=bf, labelsep=period, textfont=sl]{caption}

\newcommand{\para}[1]{{\vspace{2pt} \bf \noindent #1}}  
\newenvironment{packed_itemize}{
\begin{list}{\labelitemi}{\leftmargin=1em}
\setlength{\itemsep}{1pt}                                                           
\setlength{\parskip}{0pt}                                                                                 \setlength{\parsep}{0pt}                                                                                  \setlength{\headsep}{0pt}                                                                                 \setlength{\topskip}{0pt}                                                                                 \setlength{\topmargin}{0pt}                                                                               \setlength{\topsep}{0pt}                                                                                  \setlength{\partopsep}{0pt}                                                                               }{\end{list}}

\usepackage[strict]{changepage}

\usepackage{framed}

\definecolor{formalshade}{rgb}{0.93, 0.93, 0.93} 

\newenvironment{formal}{%
  \MakeFramed{\advance\hsize-\width\FrameRestore}%
  \noindent\hspace{-4.55pt}
  \begin{adjustwidth}{}{7pt}%
  \vspace{2pt}\vspace{2pt}%
}
{%
  \vspace{2pt}\end{adjustwidth}\endMakeFramed%
}
\definecolor{lightgreen}{HTML}{117733}
\definecolor{lred}{HTML}{FF0000}

\title{The Cool and the Cruel: \\
Separating Hard Parts of LWE Secrets}

\date{\today}

\begin{document}

\thispagestyle{empty}

\author{Niklas Nolte$^*$\inst{1} \and
Mohamed Malhou$^*$\inst{1}\inst{4} \and Emily Wenger$^*$\inst{1} \and Samuel Stevens\inst{2} \and \\ 
Cathy Li\inst{3} \and François Charton$^\dagger$ \inst{1} \and Kristin Lauter$^\dagger$\inst{1}}
\def\thefootnote{*}\footnotetext{Equal Contribution}
\def\thefootnote{$\dagger$}\footnotetext{Co-Senior authors}
\authorrunning{N. Nolte et al.}
\institute{FAIR, Meta, Menlo Park, CA, USA \email{\{nolte$^+$,mmalhou,ewenger,fcharton,klauter\}@meta.com} \and Ohio State University, Columbus, Ohio, USA  \email{stevens.994@osu.edu} \and University of Chicago, Chicago, IL, USA \email{cathyli@uchicago.edu} \and Sorbonne Université, CNRS, LIP6, F-75005 Paris, France}
\def\thefootnote{+}\footnotetext{Corresponding Author}
\maketitle              

\begin{abstract}

Sparse binary LWE secrets are under consideration for standardization for Homomorphic Encryption and its applications to private computation~\cite{cheon2017homomorphic}.
Known attacks on sparse binary LWE secrets include the sparse dual attack~\cite{Albrecht2017_sparse_binary} and the hybrid sparse dual-meet in the middle attack~\cite{Cheon_hybrid_dual}, which requires significant memory.
In this paper, we provide a new statistical attack with low memory requirement.  The attack relies on some initial lattice reduction.  The key observation is that, after lattice reduction is applied to the rows of a q-ary-like embedded random matrix $\mathbf A$, the entries with high variance are concentrated in the early columns of the extracted matrix. This allows us to separate out the ``hard part'' of the LWE secret. We can first solve the sub-problem of finding the ``cruel'' bits of the secret in the early columns, and then find the remaining ``cool'' bits in linear time.  We use statistical techniques to distinguish distributions to identify both cruel and cool bits of the secret. We recover secrets in dimensions $n=256, 512, 768, 1024$ and provide concrete attack timings. 
For the lattice reduction stage, we leverage recent improvements in lattice reduction (flatter~\cite{ryan2023fast}) applied in parallel. We also apply our new attack to RLWE with 
$2$-power cyclotomic rings, showing that these RLWE instances are 
much more vulnerable to this attack than LWE.
\end{abstract}

\section{Introduction}
\label{sec:intro}

Lattice-based cryptosystems are attractive candidates for Post-Quantum Cryptography (PQC), standardized by NIST in the 5-year PQC competition (2017-2022) and by the Homomorphic Encryption community~\cite{HES,nist2022finalists}.   
Special parameter choices, such as binary, ternary, binomial, or Gaussian secrets, small error, or small {\it sparse} secrets, are often made to improve efficiency or functionality.
Homomorphic Encryption (HE) implementations routinely assume small error and sparse binary or ternary secrets~\cite{HES,params_heean}, and the NIST-standardized Kyber schemes assume small secret and small error (binomial distribution)~\cite{nist2022finalists}.

Although these implementation choices may weaken security of the lattice-based cryptosystems, surprisingly few concrete results and benchmarks exist to quantify the actual time and resources required to attack systems with sparse secrets and small error. 
Over the last few decades, the lattice cryptography community has developed theoretical predictions of time and resources needed to attack lattice-based cryptosystems in general by studying and improving lattice reduction algorithms, such as LLL, BKZ, BKZ2.0 with various strategies such as sieving, early abort etc. (\cite{LLL,CN11_BKZ,fplll}). 
To estimate the performance of these algorithms, the community relies on the LWE Estimator~\cite{LWEestimator}, but these estimates are often inaccurate for special parameter choices and for parameter sizes that can be tested in practice. 
While some attacks on sparse secrets have been proposed, such as the Sparse dual attack~\cite{Albrecht2017_sparse_binary} and the Hybrid Sparse Dual Meet-in-the-Middle attack~\cite{Cheon_hybrid_dual}, general benchmarks for measuring attacks on small sparse secrets have not yet been developed. Existing benchmarks, such as the Darmstadt Lattice Challenges~\cite{darmstadt_lwe} are not relevant, since they do not cover such parameter choices.

\para{Our Contribution.} Given the potentially widespread adoption of lattice cryptosystems with sparse, small secrets, more work is needed to assess their hardness.  
To that end, we provide a new tool for attacking sparse small secrets, based on the observation that lattice reduction with certain parameters separates secrets into ``cruel bits'', corresponding to unreduced data which is hard to analyze, and ``cool bits'' that can be recovered in linear time once the hard bits are known. This leads us to define an attack on sparse small secrets, in three stages:

\begin{enumerate}
\item initial partial lattice reduction in parallel; 
\item brute force recovery of a small number of hard, cruel bits; 
\item statistical recovery of cool bits, and the secret.
\end{enumerate}
We provide concrete experimental results for our attack, demonstrating its efficacy on LWE instances with sparse binary secrets in dimension $n \le 768$. Furthermore, we show how this attack applies to both LWE and RLWE instances, with RLWE instances being much more vulnerable. We believe this attack should be taken into account whenever sparse secrets are being proposed, since a significant fraction of secrets are vulnerable to it (see Table~\ref{tab:experiments_recovery}), and we hope it provides a helpful practical benchmark for future research.  

\para{High-Level Attack Idea.} Our key observation is that, for an LWE instance $(\mathbf{A}\in \mathbb{Z}_q^{m \times n}, b \in \mathbb{Z}_q^m)$, lattice reduction of q-ary embedded matrices of the form 
\begin{align}
\mathbf \Lambda = 
\begin{pmatrix}
0 & q \mathbf{I}_n \\
\omega \mathbf{I}_m &\mathbf  A
\end{pmatrix},
\end{align}
produces a reduced matrix $\mathbf A'$ with a non-uniform distribution of variances over its columns; see \cref{sec:attack} for details. Specifically, the variance of the first $n_u$ columns of $\mathbf A'$ is unaffected by lattice reduction. We call the bits of the secret in the first $n_u$ columns ``cruel" (unreduced). The other $n_r = n - n_u$ columns of the matrix have significantly reduced variance, and we call the secret bits corresponding to these columns the ``cool" bits. The number of unreduced and reduced columns of $\mathbf{A}'$, $n_u$ and $n_r$, depends on the overall lattice reduction quality and the penalty parameter $\omega$, which controls a trade-off between reduction and the error variance after reduction.

Our attack leverages this observation to reduce the original LWE problem to one with higher error on only the “cruel” entries in $\bf{A}$, by essentially ignoring the ``cool" bits. If the error introduced by the ignored bits is sufficiently small, a secret guess $s^*$ where only the first $n_u$ bits are correct can be identified with sufficient statistics.
The remaining (cool) bits can then be recovered one by one.

\begin{table}[h]
\vspace{-0.85cm}
    \centering
    \small
    \caption{
        Successful LWE secret recovery attacks leveraging our cruel/cool observation for various $n$/$q$ settings. The timings do not include time for parallelized lattice reduction.}
        \vspace{0.1cm}
    \begin{tabular}{
        c
        @{\hskip 6pt}
        S[table-format=2.0]
        @{\hskip 6pt}
        S[table-format=2.0]
        @{\hskip 6pt}
        S[table-format=4.0]
    }
        \toprule
        $n$ & {$\log_2 q$} & {Hamming Weight} & {Time (Sec)} \\
        \midrule
        \multirow{2}{*}{$256$} & 12 & 12 & 3865 \\
        \cmidrule(lr){2-4}
        \multirow{2}{*}{\num{512}} & 28 & 12 & 2417\\
        & 41 & 60 & 376 \\
        \cmidrule(lr){2-4}
        \num{768} & 35 & 12 & 1291 \\ \cmidrule(lr){2-4}
        $1024$ & 50 &  17 & 6395 \\ 
        \bottomrule
    \end{tabular}
    \label{tab:highlights}
    \vspace{-0.5cm}
\end{table}

\cref{tab:highlights} presents parameter settings and timings for successful recovery of sparse binary secrets, and demonstrates the benefits of cruel-cool bit separation in non-trivial LWE settings (see~\cref{tab:experiments_recovery} for additional results).
In these examples, Hamming weights were selected so that secret recovery from reduced matrices only takes a small amount of compute. We choose those settings because the lattice reduction step is theoretically least well understood, whereas the rest of the attack can be understood better theoretically (see \cref{sec:theory}) and does not need extensive experimentation.
For example, for dimension $n=512$ and $\log_2 q =41$, our attack recovers binary secrets with Hamming weight $60$ in about $6$ minutes on one GPU.  This does not include the time spent on lattice reduction to prepare the data, which is about $12$ hours/matrix (see Table~\ref{tab:datasets}).  
Table~\ref{tab:highlights} also shows successful secret recovery for larger dimensions, such as $n=768$ and $\log_2 q =35$, for binary secrets with Hamming weight $12$, in roughly $22$ minutes using $20$ GPUs.

\para {RLWE attack.} Our attack can also be applied in the Ring-LWE setting.  When the ring is a $2$-power cyclotomic ring defined by the polynomial $x^n+1$, where $n=2^k$, then the RLWE samples corresponding to a polynomial RLWE sample, $(a(x), b(x) = a(x) \cdot s(x) + e(x))$ can be described via a skew-circulant matrix, $A_{circ}$; see Section~\ref{sec:RLWE} for details. When the matrix has this structure, we can rotate the ``cruel bits'' around (without redoing any lattice reduction) by inspecting samples from only certain indices of circulant matrices. This greatly increases our chances of recovering the secret and makes attacking the 2-power cyclotomic RLWE problem clearly easier than generic LWE.  We give concrete timings for the RLWE setting, estimate the average speed-up over LWE, and show the improved success rate in~\cref{sec:RLWE}.

\para{Outline of the paper.} \cref{sec:compare} describes related work. \cref{sec:attack} presents our attack in detail. \cref{sec:reduction} explains the lattice reduction step for processing LWE samples to use in the attack. \cref{sec:theory} contains the statistical analysis of the attack, including an estimate for number of samples needed. \cref{sec:results} presents our concrete secret recovery results and performance of comparable attacks. 
\cref{sec:RLWE} adapts the attack to the RLWE setting. 

\vspace{-0.2cm}
\section{Related Work}
\label{sec:compare}
Albrecht et al \cite{albrecht2018} classify attacks on LWE broadly into primal \cite{LP11,LN13,AFG13} and dual \cite{MR09,Albrecht2017_sparse_binary} attacks.
Some primal attacks reduce search-LWE (see \cref{sec:attack} for definition) to the unique shortest vector problem (uSVP) and solve this via lattice reduction \cite{albrecht2017revisiting}. Others employ a combination of lattice reduction and Bounded Distance Decoding (BDD) \cite{LP11}. The goal of dual attacks is to solve Decision-LWE (see \cref{sec:attack}) by reducing it to the Shortest Integer Solution (SIS) problem \cite{ajtai1996generating}, further reduced into the problem of finding short vectors in the dual lattice defined by $\{ x \in \mathbb{Z}^m_q | Ax = 0 \mod q \}$.

\para{Hybrid attacks.} The notion of hybrid attacks was introduced in 2007 by \cite{howgrave2007hybrid}, who proposed combining lattice reduction and meet-in-the-middle (MITM) attacks against NTRUEncrypt. \cite{JFRT16} extended this idea to binary-error LWE.
A collection of such hybrid attacks have been proposed for different LWE settings.
\cite{bilu2021} provide a succinct overview in their Table 1.
\cite{soncheon2019} succeeded in applying a hybrid attack to settings of sparse ternary secrets with small error via a primal attack strategy.
\cite{Albrecht2017_sparse_binary} was the first work to combine the notions of hybrid attacks in the dual lattice for sparse binary and ternary secrets and small error.
The attack benefits from optimizing a trade-off between success probability by guessing parts of the secret and performing the costly lattice reduction on the now lower-dimensional lattice. 
\cite{Cheon_hybrid_dual} improved upon \cite{Albrecht2017_sparse_binary} by using MITM for the guessing part.
\cite{espitau2020} optimized further by recognizing an inefficiency in the matrix multiplication during the guessing and considered small, non-sparse secrets.
\cite{bilu2021} extended this line of attack to arbitrary secrets and find that ``hybrid dual attacks [usually] outperform [non-hybrid] dual attacks regardless of the secret distribution''.

\para{Our work.} In contrast to (but not necessarily incompatible with) work in hybrid attacks, our attack performs partial lattice reduction on full-sized lattices arising from different subsets of available LWE samples. Our lattice reduction uses a scaling (or penalty) parameter $\omega$ similar to the one in \cite{bai_galbraith,Albrecht2017_sparse_binary}, and its utility is discussed in \cref{sec:reduction}.
Our key insight, the non-uniformity over coordinates in the reduced basis, naturally splits the problem into two parts (in this case the ``cruel'' and ``cool'' bits), reminiscent of a hybrid attack.

\begin{table}[h]
\vspace{-0.7cm}
    \centering
    \small
    \caption{Notation used in this paper.}
    \begin{tabular}{c@{\hskip 6pt}l}
        \toprule
        Symbol & Description \\
        \midrule
        $q$ & The modulus of the LWE problem considered \\
        $n$ & Problem dimension (the dimension of vectors $\mathbf a$ and $\mathbf s$) \\
        $n_u$ & The number of unreduced (aka cruel) entries in $\mathbf a$. \\
        $n_r$ &  Number of reduced (aka cool) entries in $\mathbf a$.  $n = n_u + n_r$ \\
        $\mathbf s$ & The unknown secret, used to construct $b=\mathbf a \cdot \mathbf s + e$ \\
        $\mathbf s^*$ & A candidate secret, not necessarily correct \\
        $h$ & The Hamming weight of the secret (number of 1 bits) \\         
        $h_u$ & The Hamming weight of the $n_u$ unreduced bits of the secret \\
        $h_r$ & The Hamming weight of the $n_r$ reduced bits of the secret ($h=h_u+h_r$)\\
        $\sigma_u$ & The standard deviation of unreduced entries in $\mathbf a$ (equal to $\frac{q}{\sqrt{12}}$) \\
        $\sigma_r$ & The standard deviation of reduced entries in $\mathbf a$ \\
        $\sigma_e$ & The standard deviation of error (amplified by reduction) \\
        $\sigma(x)$ & The standard deviation of the random variable $x$ \\
        $\rho$ & The reduction factor of pre-processing, i.e. the ratio $\frac {\sigma(\mathbf R \mathbf A)}{\sigma(\mathbf A)}$ \\ 
        $m$ & The number of LWE samples (reduced or unreduced) \\
        $\omega$ & The penalty used during reduction \\
        $a:b$ & integer range $[a, b)$, used for indices \\
        $\mathbf{X}_{*,i}$ & $i$th column of $\mathbf X$ \\
        \bottomrule
    \end{tabular}
    \label{tab:notation}
    \vspace{-0.3cm}
\end{table}

\section{The Attack} \label{sec:attack}
The Learning With Errors (LWE) problem, first introduced by Regev~\cite{Reg05}, can be stated in 2 forms. The {\em Search-LWE} problem is: given a random matrix \( \mathbf A \in \mathbb{Z}_q^{m \times n} \), a secret vector \( \mathbf{s} \in \mathbb{Z}_q^n \), and an error vector \( \mathbf{e} \in \mathbb{Z}_q^m \) (usually sampled from some small error distribution), find \( \mathbf{s} \), given $\mathbf A$ and 
\begin{align}
\mathbf{b} = \mathbf A\cdot \mathbf{s} + \mathbf{e} \mod q,
\end{align}
where $m$ denotes the number of LWE samples, $n$ is the lattice dimension, and $q$ is the modulus. In the related {\em Decision-LWE} problem, the objective is not to find the secret vector \( \mathbf{s} \), but to distinguish between two distributions: given a random matrix \( \mathbf A \in \mathbb{Z}_q^{m \times n} \) and a vector \( \mathbf{b} \in \mathbb{Z}_q^m \), decide whether \( \mathbf{b} \) is drawn from the distribution \( \mathbf A\cdot \mathbf{s} + \mathbf{e} \mod q \) for some secret vector \( \mathbf{s} \in \mathbb{Z}_q^n \) and error vector \( \mathbf{e} \in \mathbb{Z}_q^m \), or whether \( \mathbf{b} \) is drawn from the uniform random distribution over \( \mathbb{Z}_q^m. \) \cite{Reg05} showed that for $q$ of size polynomial in $n$, there is a reduction from Search-LWE to Decision-LWE.  See \cref{tab:notation} for notation used in the paper.

 We introduce a new attack on the LWE problem for sparse binary secrets, i.e. $\mathbf{s} \in \{0,1\}^n$, with $h$, the Hamming weight (number of $1$s in $\mathbf s$), small. The attack leverages an observation about the shape of reduced LWE matrices to separate the hard and easy parts of the secret. It works as follows. 

\para{Attack Part 1: Lattice Reduction.} We begin by applying  lattice reduction to (the rows of) an embedding $\mathbf \Lambda \in \mathbb{Z}_q^{(m+n) \times (m+n)}$ of the data:
\begin{align}
\label{eq:embedding}
\mathbf \Lambda = 
\begin{pmatrix}
0 & q \mathbf{I}_n \\
\omega \mathbf{I}_m &\mathbf  A
\end{pmatrix}
\end{align}
The reduction finds a linear transformation $\left[\mathbf{C}, \mathbf{R}\right]$ such that 
$\mathbf \Lambda^r = \left[\mathbf{C}, \mathbf{R}\right] \mathbf{\Lambda} =
\left[ \omega \mathbf{R}\mathbf{I}_m, q\mathbf{C}\mathbf{I}_n + \mathbf{RA} \right]$.
We extract the transformation matrix $\mathbf R = \mathbf \Lambda^r_{*,0:m}/\omega$,
which corresponds to the row and column operations performed to transform the matrix $A$ into its reduced form.
$\mathbf{R}$ is applied to both $\mathbf{A}$ and $\mathbf{b}$ to create reduced sample pairs.
For ease of notation, we will refer to $\mathbf R\mathbf A$ as $\mathbf A$ or ``reduced $\mathbf A$'' and $\mathbf  R \mathbf{b}$ as $\mathbf{b}$ throughout the text, and otherwise specify ``original'' pairs.

Reduction algorithms trade off the error on the reduced LWE pairs for the norm of the rows in reduced $\mathbf A$. The ``penalty parameter'' $\omega$ controls this trade-off, where higher $\omega$ causes less norm and variance reduction, but also less error amplification. Details of lattice reduction can be found in Section \ref{sec:reduction}.

\begin{figure*}[ht!]
   \vspace{-0.4cm}
    \centering
    \subfigure[]{\includegraphics[width=0.45\textwidth]{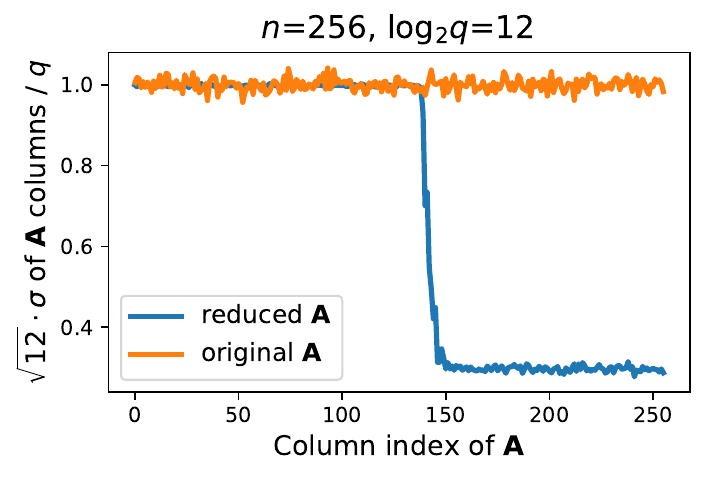}} 
    \subfigure[]{\includegraphics[width=0.45\textwidth]{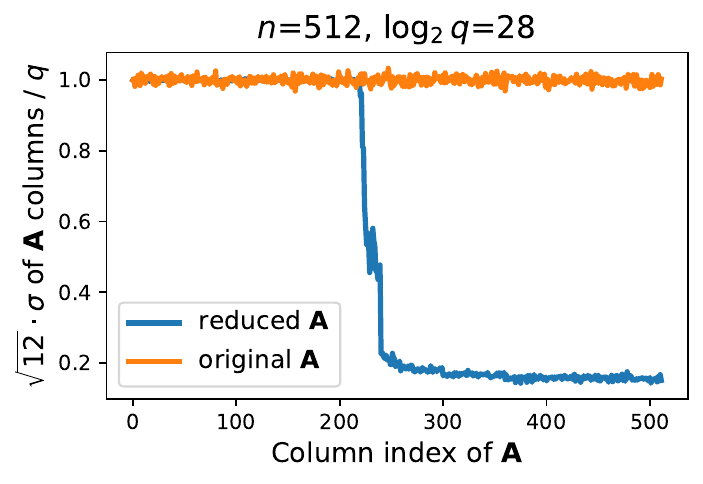}}
    \subfigure[]{\includegraphics[width=0.45\textwidth]{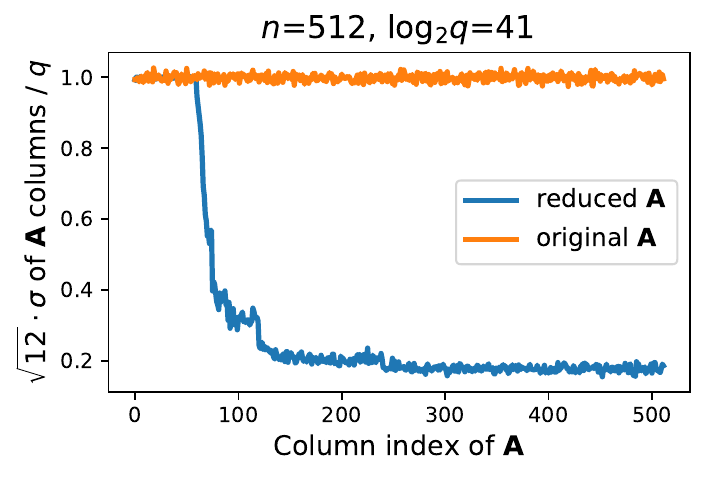}}
    \subfigure[]{\includegraphics[width=0.45\textwidth]{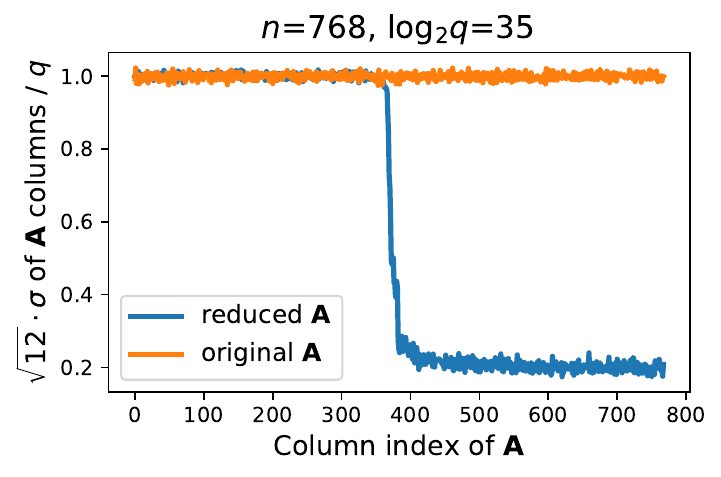}}
    \vspace{-0.2cm}
    \caption{The standard deviation of elements within each column of the $\mathbf{A}$ matrix before and after reduction (and extraction from the q-ary embedding) for various $n$/$q$ settings. The first $n_u$ unreduced columnms (left half of the figures) correspond to the ``cruel'' bits of the secret, while the remaining $n_r = n - n_u$ are the ``cool'' bits.  This phenomenon is distinct from the "z-shape" exhibited by the Gram-Schmidt orthogonalized rows of a q-ary lattice before and after lattice reduction~\cite{howgrave2007hybrid,albrecht2021lattice} (see Appendix~\ref{subsec:zshape} for details).
    }
    \label{fig:stdev_examples}
    \vspace{-0.5cm}
\end{figure*}

\para{Attack Part 2: Identify Cruel and Cool Bits.} The key observation of this attack is that the resulting reduced $\mathbf{A}$ samples all share a reduction pattern: 
\begin{formal} 
{\bf Key observation:} {\it The first $n_u$ elements of each sample vector  $\mathbf A_{*,0:n_u}$ remain unreduced and retain their uniform distribution over $\mathbb{Z}_q$,
and all subsequent $n_r = n - n_u$ elements of $\mathbf A_{*, n_u:n}$ are heavily reduced. }
\end{formal}

To illustrate this key observation, we plot the standard deviation over columns in $\mathbf A$ against the column index in \cref{fig:stdev_examples} for various $n$ and $\log_2 q$ settings. 
This observation of cruel and cool bits inspires the following hypothesis:
 
\begin{formal}
{\bf Hypothesis:} For any row $\mathbf a$ of reduced matrix $\mathbf A$, the $n_r$ last entries contribute comparatively little to the overall dot product $\mathbf a \cdot \mathbf s$. Thus, for a sparse enough secret, the Decision-LWE may be solvable without correctly guessing the $n_r$ secret bits.\end{formal}
In other words, this shape resulting from lattice reduction means we only need to solve LWE on a smaller dimension $n_u < n$. In essence, \textit{we separate out the hard part of the LWE secret at the cost of increased error}. We validate this hypothesis in the remainder of this paper. 

\begin{figure}[h]
    \centering
    \includegraphics[width=0.65\linewidth]{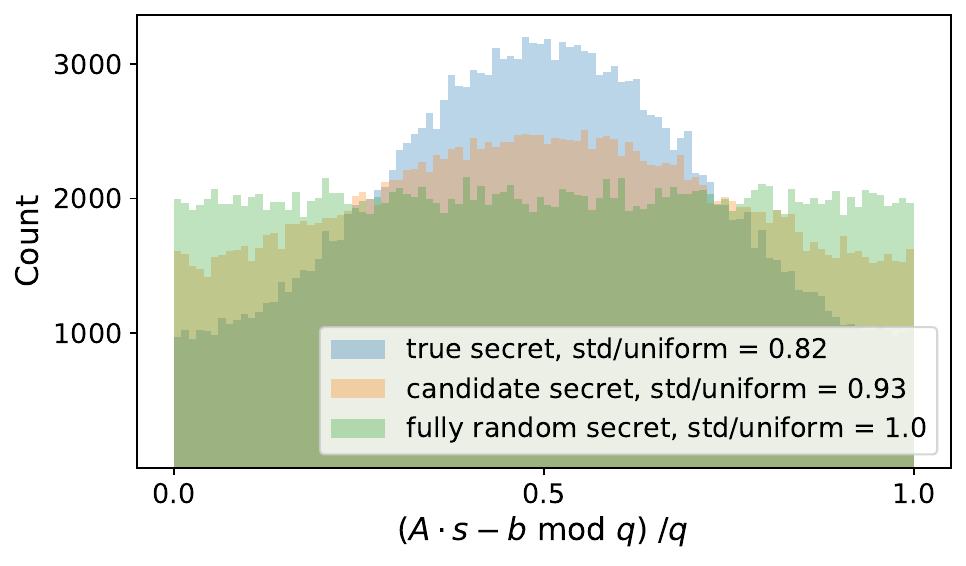}
    \caption{Histogram of 4 million samples of the residual $(\mathbf a \cdot \mathbf{s^*} - b) \mod q$ (centered and normalized) for reduced data and different secret guesses $\mathbf{s^*}$ ($n=512$, $\log_2 q = 41$, $h=20$). The three histograms correspond to three different guesses. Blue corresponds to the true secret, orange to the secret candidate where the first $n_u=75$ entries are guessed correctly, and green to a random secret with $h=20$. The numbers in the legend correspond to sample standard deviation relative to uniform standard deviation $\frac{q}{\sqrt{12}}$.
    }
    \label{fig:as-bmodq}
    \vspace{-0.3cm}
\end{figure}

\para{Attack Part 3: Secret Guessing.} Next, we leverage statistical properties of the cool and cruel regions to guess the secret. Figure \ref{fig:as-bmodq} shows a histogram of the distribution of the residuals of reduced data $\mathbf a\cdot \mathbf{s} - b \mod q$ (in blue) and $\mathbf a\cdot \mathbf{s}^* - b \mod q$ (in orange), where $\mathbf{s}^*$ shares the first $n_u$ entries with $\mathbf{s}$
and has all other $n_r$ elements set to $0$. Green corresponds to a random guess for the secret with equal hamming weight. The orange and the green distributions are distinct, enabling us to detect the correctness of the first $n_u$ bits in the secret without knowing the others. Given the size of the error on the reduced LWE problem, so far we have not found a more efficient attack for guessing the first $n_u$ secret bits than brute force. One could think of using one of the hybrid attack strategies outlined in the related work on the cruel part, but we leave this for future work.

Given a secret guess $\mathbf{s}^*$ with correct $n_u$ cruel bits, the $n_r$ remaining cool bits can be recovered easily, given enough samples. One fast method is greedy recovery, presented in Algorithm \ref{alg:greedy_cool}. It works as follows: 
start with a candidate secret 
$\mathbf{s}^*$ which shares the first $n_u$ cruel bits with $\mathbf{s}$
and has all other $n_r$ bits set to $0$.
Starting with the first cool bit, flip that bit of $\mathbf{s}^*$ to $1$. Then calculate if this flip increases the 
standard deviation of the resulting 
$\mathbf a \cdot \mathbf{s}^* - {b} \mod q$ (over $m$ reduced samples $(\mathbf a,b)$). If so, flip it back to $0$ and move to the next bit. Note the standard deviation here is a practical proxy for a proper statistical test. Section \ref{sec:practical} expands on this. Intuitively, greedy recovery works because every correct bit flipped to 1 will reduce the orange distribution in Figure \ref{fig:as-bmodq} closer to the blue and every wrong one will introduce more noise, sending it closer to the green. 
\begin{algorithm}
\caption{Greedy recovery of cool bits}\label{alg:greedy_cool}
\begin{algorithmic}[1]
\For{ $i \in n - n_r : n$}
    \State $ \mathbf{s}^*[i] \gets 0 $
    \State $ \sigma_0 \gets \sigma(\mathbf a \cdot \mathbf{s}^* - {b} \mod q)$ 
    \State $ \mathbf{s}^*[i] \gets 1 $
    \State $ \sigma_1 \gets \sigma(\mathbf a \cdot \mathbf{s}^* - {b} \mod q)$ 
    \If{$ \sigma_0 \leq \sigma_1 $}
        \State $ \mathbf{s}^*[i] \gets 0 $
    \Else
        \State $ \mathbf{s}^*[i] \gets 1 $
    \EndIf
\EndFor
\end{algorithmic}
\end{algorithm}
\vspace{-0.5cm}

Otherwise, one can employ more principled gradient-free optimization methods, like simulated annealing. Annealing works better with limited data, but is slower. The best choice depends on the setting.
All methods work efficiently due to the reduced variance of the corresponding entries in $\mathbf A$.

\subsection{Practical Considerations} \label{sec:practical}
\vspace{-0.1cm}

\para{Implementation.} As described in detail in Section \ref{sec:reduction}, we sample views of the $4\cdot n$ sized lattice defined by the LWE samples and reduce them individually, forming a large set of reduced LWE samples for use in the attack, following~\cite{picante,verde}. After identifying the cool and cruel bits, secret recovery runs in two parts: first, guessing the cruel secret bits on the the reduced LWE problem with large error, then the recovery of the full secret via Algorithm \ref{alg:greedy_cool}.

We use brute force enumeration for cruel secret bit recovery. All code is implemented in Python, and we employ GPU acceleration via \texttt{pytorch} \cite{pytorch}.
Batches of secret candidates are continuously sent to the device and evaluated on a suitable amount of data to distinguish the residuals $\mathbf A\cdot \mathbf{s}^* - \mathbf{b} \mod q$ if the true (part of the) secret was found, see the orange histogram in Figure \ref{fig:as-bmodq}.
Since our brute force is perfectly parallelizable, we can distribute the work to an arbitrary number of GPUs or CPUs. Candidate secrets are enumerated in ascending Hamming weight, and a record of the top-$k$ candidates, according to a metric defined below, are evaluated at constant intervals throughout the enumeration. Evaluation involves running Algorithm \ref{alg:greedy_cool} and then checking if the resulting secret produces nearly correct ($\mathbf{A, b}$) pairs.
The evaluation frequency can be traded off against the amount of data used during enumeration and $k$, the number of candidates kept in the top-$k$ set.

\para{Candidate Evaluation.} The metric by which candidate secrets are sorted can involve any test that distinguishes from uniformity in the residual, but in practice it must be fast. Most non-parametric tests like Kolmogorov-Smirnov or Kuiper involve calculating a Cumulative Distribution Function (CDF) and thus sort the data, which is slow.
A likelihood ratio test is generally more powerful, but evaluating such a test on a distribution modulo $q$ involves evaluation over multiple ``wraps'' around the modulus.
Under certain assumptions, detailed in Section \ref{sec:theory}, the distribution we aim to distinguish from uniform is a mod-Gaussian involving an infinite sum:
\begin{align}
f(x) \propto \sum_{k=-\infty}^{\infty} \mathcal{N}(x ; \mu + k\cdot q, \sigma^2).
\end{align}
Even with suitable truncation, the runtime of a likelihood ratio test will at best be $O(N\cdot K)$ with $N$ samples and truncation to $K$ elements of the sum.
\cite{espitau2020} also explores a way of specifically distinguishing modular Gaussians via a periodic aggregation metric $Y=\sum^N_{i=1} \exp(2\pi iX_i)$, see their Algorithm 2.
We thought, since both distributions, uniform and mod-Gaussian are fully defined by their first two moments and have their mean fixed to the same value, a direct comparison of the second moment suggests itself as a simple but powerful distinguishing method.
Thus, we choose the direct variance comparison described in Algorithm~\ref{alg:greedy_cool} due to its simplicity and computational efficiency. Empirically, we found that it was similarly powerful to a Kuiper test, and more than 5x faster in the settings considered.

\para{Other Performance Considerations.}
Values of $\mathbf A, \mathbf b$ are of the order of $q$ and thus large in the considered scenarios. However, high precision is not needed during the enumeration and evaluation, since all precision is drowned out by the large error. Since modern GPUs work best with 16-bit floating point numbers, we scale all values down to a range appropriate for those representations. We also employ \texttt{pytorch}  compilation to alleviate some of the overhead of python being an interpreted language. This feature compiles, optimizes and fuses tensor operations, giving a 2-3x speed improvement over the default eager execution.

Our implementation can evaluate roughly 5 billion secrets of dimension $n=512$ through an NVIDIA V100 GPU in about 1 hour during the brute force search. The greedy secret recovery step is fast and runs infrequently. Additional speedups can be had with additional optimizations, such as the use of dedicated CUDA kernels. Additionally, our attack does not yet exploit the recursive search space structure trick from \cite{espitau2020}.

\section{Lattice Reduction}
\label{sec:reduction}

Here, we briefly describe the lattice reduction techniques used to produce the data in our attacks. Classical lattice reduction algorithms, such as LLL, BKZ, BKZ2.0 (\cite{LLL},\cite{CN11_BKZ},\cite{fplll})
seek to reduce the size of vectors in a lattice. 
Attacks on LWE such as the uSVP and dual attacks use 
lattice reduction algorithms to find the shortest vector, or a {\it short enough} vector, in a specifically constructed lattice.
Our attack leverages lattice reduction methods, but instead of attempting to find a single very short vector, we 
use reduced lattices as data for a distinguishing attack.
For our attack, we apply lattice reduction techniques to the embedded matrix in \cref{eq:embedding}.

Lattice reduction can be viewed as finding an integer-valued transformation matrix $\mathbf R$ which, when multiplied by $\mathbf A$ modulo $q$, reduces the Frobenius norm of $\mathbf A$. Given LWE samples in the form of a matrix $\mathbf A$ and a vector $\mathbf b =\mathbf A \cdot \mathbf s + \mathbf e$, we can use lattice reduction to reduce the entries of the matrix $\mathbf A$, yielding $\mathbf R\mathbf A$.  Then the corresponding LWE samples are ($\mathbf R\mathbf A$, $\mathbf R\mathbf b = \mathbf R\mathbf A \cdot \mathbf s + \mathbf R\mathbf e$).
Because $\mathbf R$ is applied to the noisy vector $\mathbf b$ as well, the initial noise in $\mathbf b$ is increased by the magnitude of the elements in $\mathbf R$.

For this attack, it is not always beneficial to reduce as much as possible, but rather to trade off reduction quality with the magnitude of the error $\mathbf R\mathbf e$ introduced by the reduction via the parameter $\omega$. Intuitively, the stronger the reduction, the higher the error.  The error magnitude determines the width of the blue and orange distributions in \cref{fig:as-bmodq}. \cref{sec:theory} expands on the theory which describes this trade-off.

\subsection{Our Reduction Methods}

\label{sec:preprocess}
\para{Generating sufficient LWE data.} Our attack relies on the ability to distinguish distributions from uniform, and successful distinguishing requires sufficient data. Details are discussed in \cref{sec:theory} below. An attacker might not have access to enough samples to run our attack in a real world scenario. However, sub-sampling comes to the rescue~\cite{picante}. From an initial number of $m_0$ samples, one can select a subset of $m_1$ samples, create a new $\mathbf A$ matrix, embed it in $\Lambda$, and perform lattice reduction.
This can be done many times, up to ${m_0 \choose m_1}$ times. This technique ``inflates'' the number of samples usable for this attack dramatically, at the cost of reduction of many matrices corresponding to different sub-lattices.

The SALSA Picante~\cite{picante} attack, which proposed this sub-sampling method,  eavesdrops $4n$ LWE samples, then sub-samples $n$ samples at a time and applies lattice reduction to $2n \times 2n$ q-ary matrices. 
Instead, here we sub-sample $m = \text{round}(0.875 n)$ from $4n$ LWE samples.
We then form q-ary matrices of size $(m+n) \times (m+n)$ and reduce them.

\para{Reduction algorithms.}  We leverage 3 different algorithms in our lattice reduction step: \texttt{BKZ2.0}~\cite{CN11_BKZ}, \texttt{flatter}~\cite{ryan2023fast}, and \texttt{polish}~\cite{charton2023efficient}. Each algorithm has strengths and weaknesses, and combining them maximizes the strengths of each. \texttt{BKZ2.0} provides strong reduction, but is slow for large block size and high dimensional lattices. In contrast, \texttt{flatter} is fast and can reduce lattices with $n\ge 512$ and large $q$, but only has reduction performance analogous to $\texttt{LLL}$. The \texttt{polish} algorithm runs at the end of a \texttt{flatter} or \texttt{BKZ2.0} loop~\cite{verde} and ``polishes'' by iteratively orthogonalizing matrices. It provably produces strictly decreasing vector norms when run, is implemented in C, and runs very fast.

\para{Measuring performance.} We measure lattice reduction performance via $\rho$, the ratio of the standard deviation of the entries of the reduced lattice to the expected standard deviation of the entries of a random lattice: 
\begin{align}
\rho =  \frac{\sigma(\mathbf A)}{\sigma(\mathbf A_{\text{initial}})} = \frac{\sigma(\mathbf A)}{\frac{q}{\sqrt{12}}},
\end{align}
with $\sigma(\mathbf A)$ the standard deviation of all entries of $\mathbf A$, and $\sigma(\mathbf A_{\text{initial}})$ the unreduced, original LWE sample matrix that follows a uniform distribution.
When $\rho = 1$, lattice reduction did not succeed. When $\rho \ll 1$, a significant reduction in lattice norms has been achieved. Our attack depends crucially on $\rho$, since it determines the number of cruel bits in the secret.

\para{Interleaved reduction strategy.} Empirically, we found that combining these three algorithms gave the best reduction performance, in terms of both time and quality. \texttt{flatter} sometimes gets ``stuck'' and cannot further reduce a given lattice, particularly for smaller $q$.  In these instances, ``interleaving'' \texttt{BKZ2.0} and \texttt{flatter} provided a higher quality reduction while helping \texttt{flatter} jump out of these minima (see Table~\ref{tab:flatter_performance} for a comparison of different ways of combining \texttt{flatter} and \texttt{BKZ2.0}).  Our interleaving algorithm starts each reduction run with several rounds of \texttt{flatter} and declares a ``stall'' when \texttt{flatter} runs $3$ times  but produces an average reduction in $\rho$ of less than $0.001$. Then, the reduction runs \texttt{BKZ2.0}, and when \texttt{BKZ2.0}  stalls, it switches back to \texttt{flatter} and repeats until the reduction converges. \texttt{polish} runs after each \texttt{BKZ2.0} or \texttt{flatter} step.

\begin{table}[h]
\vspace{-0.5cm}
\centering
\small
\caption{ 
    Comparing performance of $3$ preprocessing methods for $n=512$ and varying $\log_2 q$: \texttt{flatter} only, \texttt{flatter}-initialized BKZ (e.g. \texttt{flatter} $\rightarrow$ \texttt{BKZ}) and Interleaving. Each entry in the table reports $\rho$/hours. 
    \textbf{Bold} indicates most reduction $\rho$ with ties broken by shortest time.
    }
\resizebox{0.99\textwidth}{!}{
\begin{tabular}{ccccccccccc}
\toprule
\multicolumn{2}{c}{$\log_2 q$} & 20 & 22 & 24 & 26 & 28 & 30 & 32 & 35 & 38 \\ \midrule
\multirow{3}{*}{$\omega=4$} & Interleave & {\bf 0.77/12} & {\bf 0.75/12} & 0.72/14 & {\bf 0.69/10} & {\bf 0.67/14} & {\bf 0.64/24} & {\bf 0.57/52}& {\bf 0.50/55} &{\bf 0.44/56} \\
 & \texttt{flatter}$\rightarrow$\texttt{BKZ} & 0.81/92 & 0.79/92 & 0.76/ 92 & 0.75/114 & 0.97/72 & 0.65/109 & 0.63/55 & 0.60/63 & 0.58/60 \\
 & \texttt{flatter} only & 0.97/72 & 0.97/72 & {\bf 0.72/2} & 0.7/3 & 0.68/4 & 0.68/5 & 0.58/54 & 0.58/132 & 0.46/94 \\ \midrule
\multirow{3}{*}{$\omega=10$} & Interleave & {\bf 0.80/11} & {\bf 0.77/14} & {\bf 0.74/14} & {\bf 0.71/14} & {\bf 0.68/13} & {\bf 0.66/13} & {\bf 0.59/53} & {\bf 0.53/54} & 0.47/57 \\
 & \texttt{flatter}$\rightarrow$\texttt{BKZ} & 0.81/93 & 0.80/93 & 0.78/93 & 0.76/114 & 0.76/92 & 0.97/72 & 0.65/50 & 0.62/55 & 0.58/58 \\
 & \texttt{flatter} only & 0.97/72 & 0.97/72 & 0.97/72 & 0.72/2 & 0.69/3 & 0.68/7 & 0.62/74 & 0.58/130 & {\bf 0.46/98} \\ \bottomrule
\end{tabular}
}
\label{tab:flatter_performance}
\vspace{-0.3cm}
\end{table}

\para{Early termination.} To increase the number of reduced LWE matrices produced from a single lattice reduction run, we leverage an early termination strategy. When the standard deviation reduction ratio $\rho$ stalls , we export the reduced matrix, then sub-sample a new $(m+n) \times (m+n)$ q-ary matrix from the initial set of $n$ LWE samples and start again. This better leverages available compute. 

\subsection{Final Datasets} 
We apply the interleaved reduction strategy described above on the $5$ different parameter sets. 
The number of cruel bits, $n_u$, is strongly correlated with the reduction factor $\rho$: better reduction (lower $\rho$) yields fewer cruel bits.
Because brute-forcing the $n_u$ bits is the bottleneck for our attack, we optimize the lattice reduction configuration to maximize reduction.
Reduction is trivially parallelizable, since each matrix can be processed on a different core, but is slower for larger dimensions $n$ and moduli $q$.

\begin{table}[h]
\vspace{-0.4cm}
    \centering
    \small
    \caption{
        Datasets used in concrete attacks.
        $\beta_1$ and $\beta_2$ are block sizes used in lattice reduction. 
        $\sigma_{e}$ is the ratio of standard deviation mod $q$ of the magnified error $\mathbf R\mathbf e$ to standard deviation of uniform distribution mod $q$.
        $n_u$ is the number of cruel bits. 
        We define any bits with standard deviation greater than $\frac{1}{2}\sigma(A_{\text{initial}})$ as cruel bits.
    }
    \vspace{0.1cm}
    \begin{tabular}{
        c
        c
        c
        @{\hskip 12pt}
        S[table-format=1.3]
        S[table-format=2.1]
        S[table-format=1.3]
        @{\hskip 6pt}
        S[table-format=3.0]
        @{\hskip 6pt}
        S[table-format=3.1]
    }
        \toprule
        \multicolumn{3}{c}{Inputs} & \multicolumn{5}{c}{Outputs} \\
        \cmidrule(lr){1-3} \cmidrule(lr){4-8}
        {$n$} & {$\log_2q$} & $\beta_1, \beta_2$ & {$\rho$} & {Hrs/Matrix} & {$\sigma_e$} & {$n_u$} & {$\rho^2 n$}\\
        \midrule
        \multirow{1}{*}{256} & \multirow{1}{*}{12} & \multirow{1}{*}{$35, 40$} & 0.769 & 15 & 0.952 & 143 & 151.3 \\
        \multirow{1}{*}{512} & 28 & \multirow{1}{*}{$18, 22$} & 0.677 & 12 & 0.692 & 228 & 234.6 \\
        512 & 41 & $18, 22$ & 0.413 & 13 & 0.337 & 75 & 87.3 \\
        768 & 35 & $18, 22$ & 0.710 & 15 & 0.938 & 373 & 387.1 \\
        1024 & 50  & $18,22$ & 0.704 & 36 & 0.827 & 495 & 506.1 \\
        \bottomrule
    \end{tabular}
    \vspace{-0.4cm}
    \label{tab:datasets}
\end{table}

\subsection{Quality of Reduction and Our Attack}

Experimentally we observe that the number of cruel bits, $n_u$, is roughly $\rho^2 n$.  The data sets presented in Table~\ref{tab:datasets} confirm this observation.
Intuitively, this makes sense when assuming column independence and that the standard deviation of the cool bits $\sigma_r$ is much smaller than that of the cruel bits $\sigma_u$:
\begin{align}
\sigma^2_{\text{total}} &= \frac{1}{n}(n_u \cdot \sigma_u^2 + n_r \cdot \sigma_r^2) \approx \frac{1}{n}(n_u \cdot \sigma_u^2)
\end{align}
During our experiments we usually find $\sigma_r < 0.3 \sigma_u$.

\section{Statistical Tests and Analysis}
\label{sec:theory}

Three statistical tests must be performed during the attack: testing whether a brute force guess of the cruel bits is correct, testing whether cool bits are zero or one (in Algorithm~\ref{alg:greedy_cool}), and testing a full secret guess. The first two tests are performed on reduced LWE samples, obtained during the pre-processing phase. The third involves a simple check of residuals on the original LWE sample and will not be discussed here. We elaborate on the statistical properties of these tests, estimate the number of samples needed, and illustrate our results with concrete attack statistics.

\subsection{Testing Brute-forced Guesses of the Cruel Bits} 

Suppose a guess of the cruel bits was made, and let $\mathbf s^*$ be the secret guess derived by setting all cruel bits to their guessed values and all cool bits to zero. We are given $m$ reduced LWE samples $(\mathbf a,b)$, with the same secret $\mathbf s$, error $e$ and reduction factor $\rho$. We consider the random variable $x = \mathbf a \cdot \mathbf s^* - b= \mathbf a \cdot (\mathbf s^* - \mathbf s)-e \mod q,$ centered to $(-\frac q2,\frac q2)$. Our intuition from Section \ref{sec:attack} is that the standard deviation of $x$ (calculated on $m$ reduced samples) will be lowest when all cruel bits are correctly predicted. 

More precisely, let $n$ be the problem dimension and $n_u$ the number of cruel bits. Over reduced LWE samples $(\mathbf a,b)$, the $n_u$ first coordinates of $\mathbf a$, which correspond to the cruel bits, have standard deviation $\sigma_u=\frac {q}{\sqrt{12}}$ (the standard deviation of the uniform distribution). 
The standard deviation $\sigma_r$ of the $n_r$ remaining coordinates of $\mathbf a$ (the cool bits) can be derived from the reduction factor $\rho$ and $n_u$ under the assumption that all coordinates of $\mathbf a$ are independent after reduction, as follows:
\begin{align}
\rho^2 \sigma_u^2 = \frac{n_u}{n} \sigma_u^2+\frac{n_r}{n}\sigma_r^2,
\end{align}
and therefore 
\begin{align}
\sigma_r = \sqrt{\frac{\rho^2 n - n_u}{n_r}} \sigma_u.
\end{align}

\noindent Applying these formulas to the examples in Table~\ref{tab:datasets}, we have:
\begin{packed_itemize}
\item $\sigma_r=0.27 \sigma_u$ for $n=256$ and $\log_2 q=12$ ($\rho=0.769$ and $n_u=143$)
\item $\sigma_r=0.15 \sigma_u$ for $n=512$ and $\log_2 q=28$ ($\rho=0.677$ and $n_u=228$)
\item $\sigma_r=0.17 \sigma_u$ for $n=512$ and $\log_2 q=41$ ($\rho=0.413$ and $n_u=75$)
\item $\sigma_r=0.19 \sigma_u$ for $n=768$ and $\log_2 q=35$ ($\rho=0.710$ and $n_u=373$)
\item $\sigma_r=0.14 \sigma_u$ for $n=1024$ and $\log_2 q= 50$ ($\rho = 0.703$ and $n_u = 495$)
\end{packed_itemize}

When all cruel bits are correctly predicted, and if $h_r$ is the number of ones in the $n_r$ cool bits of the secret (a known value since the Hamming weight of the secret is known), the variance of $x$ is $\sigma^2_{\mathcal{G}_q}(x)=F_q(h_r\sigma_r^2 + \sigma_e^2)$,
with $F_q(v)$ the variance of a centered Gaussian modulo $q$ (which we call $\mathcal{G}_q$) with variance $v$ and a modular operation. We assume independence in the reduced samples and therefore a Gaussian distribution of $x$ before the modulo operation, 
so the Central Limit Theorem (CLT) holds.

When some cruel bits are incorrect, the residual distribution becomes almost indistinguishable from uniform, because even one incorrect cruel bit significantly increases the Gaussian variance.

\para{Samples needed for cruel bit recovery.} We now derive $M$, the number of reduced samples needed to verify cruel bit guesses, and consequently the amount of pre-processing resources needed for our attack. Since in practice the distribution of residuals is indistinguishable from uniform over $[0,q]$ when the cruel bits are incorrectly chosen, we frame the task of identifying correct cruel bits as distinguishing from uniform. As describe in Section \ref{sec:practical}, we do so by measuring the variance.
First, we calculate $\hat{\sigma}^2(x)$ over $M$ reduced samples via
\begin{align}
\hat{\sigma}^2 = \frac{1}{M}\sum_{i=0}^{M-1}x_i^2.
\end{align}
We can construct a lower confidence bound $\iota$ for the variance under the null hypothesis of uniformity, assuming a Gaussian distribution of $\hat{\sigma}^2$ under the CLT:
\begin{align}
\iota(\alpha, M) = \sigma_{\mathcal{U}_q}^2 + G^{-1}(\alpha) \cdot \sqrt{\frac{\sigma^2_{\mathcal{U}_q^2}}{M}} =  \sigma_{\mathcal{U}_q}^2 + G^{-1}(\alpha) \cdot \sqrt{\frac{\sigma_{\mathcal{U}_q}^2}{M} \cdot \frac{q^2-4}{15}},
\end{align}
where $G^{-1}$ is the percent point function of $\mathcal{N}(0,1)$ and $\sigma_{\mathcal{U}_q}^2$ (resp. $\sigma^2_{\mathcal{U}_q^2}$) is the distribution variance of $x$ (resp. $x^2$) under the null hypothesis of uniformity. $\alpha$ is the false negative error, i.e. the probability of rejecting a true null hypothesis. Since we are brute-forcing a large number of possible solutions, the false negative error $\alpha$ should be small (i.e. $\alpha <$ the inverse of the number of secret candidates).

Given $\alpha$ and $\beta$ false positive error (the probability of failing to reject a false null hypothesis), we can estimate the minimum number of samples needed for our attack to succeed with probability $\approx 1-\beta$ using CLT (see \ref{sec:m_formula_dx} for more details): 
\begin{align}
 \label{eq:Msamples}
 M(\alpha, \beta) =  \Biggl [\frac{ G^{-1}(\alpha) \sigma_{\mathcal{U}_q^2} + G^{-1}(\beta)\sigma_{\mathcal{G}_q^2}}{(\sigma^2_{\mathcal{G}_q} - \sigma^2_{\mathcal{U}_q})} \Biggr ]^2 
\end{align}
where $ \sigma_{\mathcal{G}_q}$ resp. $ \sigma_{\mathcal{G}_q^2}$ is the standard deviation of 
$x$ resp. $x^2$ when $x\sim \mathcal{G}_q$ i.e. discrete centered Gaussian mod $q$. Table~\ref{tab:datasummary} gives concrete calculations of $M$ for various $h$ secrets.

\vspace{-0.3cm}
\begin{table}[h]
\vspace{-0.5cm}
    \small
    \centering
    \caption{\small Number of reduced samples $M$ needed in the statistical test for false negative error $\alpha = 2^{-128}$ and false positive error $\beta = 10^{-5}$ for LWE settings specified by $n$, $\log_2 q$, $n_u$, and secret Hamming weight $h$. The Worst Case number of samples corresponds to the case where all the 1s in the secret are in the reduced region $h_r = h$, whereas the Average case assumes the average number of 1s in that region, which is the fraction $\frac{n_r}{n}h$.
    Because overall lattice reduction depends on both $n$ and $\log_2 q$, problem difficulty, and thus $M$, does not monotonically increase with respect to only $n$.
    } 
    
    \vspace{0.1cm}
    \label{tab:datasummary}
    \setlength{\tabcolsep}{5pt}
    \begin{tabular}{cccccc}
    \toprule
     $n$ & $\log_2 q$ & $n_u$ & $h$ & Worst case $M$ & Average case $M$ \\
     & & & &$(h_r = h)$ &$(h_r = \frac{n_r}{n}h)$ \\ \midrule
    256 & 12 & 143 & 12 & $5.67\times 10^{4}$ & $1.12\times 10^{4}$ \\ 
512 & 28 & 228 & 20 & $3.29\times 10^{3}$ & $1.66\times 10^{3}$ \\ 
512 & 41 & 75 & 60 & $1.98\times 10^{4}$ & $1.02\times 10^{4}$ \\ 
768 & 35 & 373 & 20 & $3.13\times 10^{4}$ & $9.74\times 10^{3}$ \\ 
768 & 35 & 373 & 64 & $6.21\times 10^{6}$ & $1.47\times 10^{5}$ \\ 
1024 & 50 & 495 & 15 & $2.37\times 10^{3}$ & $1.35\times 10^{3}$ \\ 
    \bottomrule
    \end{tabular}
    \vspace{-0.2cm}
\end{table}

These results indicate that in the easiest settings (e.g. $n=512$) a few thousand reduced samples are enough to test cruel bit guesses. For $n=256$, a few tens of thousands of reduced pairs are needed.
As indicated in Section \ref{sec:practical}, the amount of data used can be traded off by running the cool bit estimation and subsequent secret check more often.

\vspace{-0.5cm}

\subsection{Testing Cool Bit Predictions.}\label{sec:cool_bits}

Once the cruel bits have been guessed, we can use Algorithm~\ref{alg:greedy_cool} to recover the cool bits of the secret one bit at a time. As before, we will estimate the standard deviation of $x=\mathbf a \cdot \mathbf s^* - b$ over a sample of reduced LWE pairs, but the cruel bits in our guess $\mathbf s^*$ are now assumed to be correct. For each cool coordinate, $k$, we compare two guesses, $\mathbf s_0^*$ and $\mathbf s_1^*$, which agree with the secret on the $k-1$ first characters, are zero on the $n-k$ characters, and have their $k$-th bit set to $0$ and $1$ respectively. If the $k$-th bit is zero, $x_0=\mathbf a\cdot \mathbf s_0^* - b$ should have a lower standard deviation than $x_1=\mathbf a\cdot \mathbf s_1^* - b$. 

Suppose that all the cruel bits and the $k-1$ first cool coordinates have been correctly guessed, and there are $h^*$ one bits to be discovered in the remaining cool coordinates. Let the null hypothesis be $H_0: \mathbf s_k = 0$ and the alternative hypothesis: $H_1: \mathbf s_k = 1$. The residual $x_0=\mathbf a\cdot \mathbf s_0^* - b$ has in both cases variance 
$\sigma^2_0=F_q(h^*\sigma_r^2+\sigma_e^2)$.
Under $H_0$ the variance of $x_1=\mathbf a\cdot \mathbf s_1^* - b$ will be
$\sigma^2_{+1}=F_q((h^*+1)\sigma_r^2+\sigma_e^2) > \sigma^2_0$.
Under $H_1$ the variance of $x_1$ is 
$\sigma^2_{-1}=F_q((h^*-1)\sigma_r^2+\sigma_e^2) < \sigma^2_0$.

In each iteration $k$ of \cref{alg:greedy_cool}, our objective is to determine whether the difference in estimated variance, denoted as $\delta (\sigma^2) := \hat{\sigma}^2(x_{1}) - \hat{\sigma}^2(x_0)$, is more closely aligned with $\sigma^2_{+1} - \sigma^2_0 > 0$ or $\sigma^2_{-1} - \sigma^2_{0} < 0$. To achieve this, we estimate the variance difference using enough samples. Without the need for optimal threshold tuning, we reject the null hypothesis if the variance difference $\delta(\sigma^2)$ is negative, and accept it otherwise.

\begin{table}[h]
\vspace{-0.3cm}
    \centering
    \small
    \caption{
        {\bf Concrete attacks on different secrets.}  
        $n = $ dimension, $\log_2 q = $ modulus size, 
        $h = $ Hamming weight of the secret. 
        $h_u = $ \# cruel bits equal to $1$. 
        \% and cumulative \% show the percentage of hamming weight $h$ secrets which have $h_u$ cruel bits (or fewer).
        All attacks run on 1 GPU, except the toughest settings on 20 GPUs in parallel.
    }
    \vspace{0.1cm}
    \begin{tabular}{ccccccc}
        \toprule
        $n$ & $\log_2 q$ \quad & $h$ & $h_u$ & \quad \% (cum. \%) \quad & Samples Used & \quad Time (1 V100 GPU) \\
        \midrule
        \multirow{4}{*}{256} & \multirow{4}{*}{12} & 12 & 4 & 6.7 (9.5) & \multirow{4}{*}{200K} & 28\,s \\
        & & 12 & 5 & 14.1 (23.6) & & 84\,s, 241\,s \\
        & & 12 & 6 & 21.2 (44.8) & & 3865\,s, 4098\,s \\
        & & 12 & 7 & 23.0 (67.8) & & 23229\,s, 26229\,s \\
        \midrule
         \multirow{3}{*}{512} & \multirow{3}{*}{28} & 12 & 3 & 9.5 (13.9) & \multirow{3}{*}{200K} & 29\,s \\
        & & 12 & 4 & 17.5 (31.4) & & 70\,s \\
        & & 12 & 5 & 22.7 (54.0) & & 2417\,s, 3510\,s \\ 
        \midrule
        \multirow{2}{*}{512} & \multirow{2}{*}{41} & 60 & 7 & 16.6 (53.6) & \multirow{2}{*}{1M} & 376\,s, 341\,s \\
        & & 60 & 8 & 15.5 (69.1) & & 1555\,s \\
        \midrule
        \multirow{4}{*}{768} & \multirow{4}{*}{35} & 10 & 3 & 13.1 (19.5) & \multirow{4}{*}{200K} & 165\,s, 168\,s \\
        & & 10 & 4 & 21.7 (41.2) & & 269\,s, 745\,s \\
        \cmidrule(lr){3-4}
        & & 12 & 4 & 13.5 (22.1) & & 607\,s, 688\,s \\
        & & 12 & 5 & 20.5 (42.6) & & 1291\,s (scaled to 20 GPUs) \\
        \midrule
        \multirow{4}{*}{1024} & \multirow{4}{*}{50} & 10 & 3 & 13.2 (20.0) & \multirow{4}{*}{100K} & 59\,s, 64\,s \\
        \cmidrule(lr){3-4}
        & & 13 & 4 & 10.2 (16.0) & & 1304\,s \\
        \cmidrule(lr){3-4}
        & & 17 & 5 & 5.8 (9.1) & & 6395\,s (scaled to 20 GPUs) \\
        \bottomrule
    \end{tabular}
    \vspace{-0.4cm}
    \label{tab:experiments_recovery}
\end{table}

\section{Recovery Results}
\label{sec:results}

In this section we provide concrete performance results for our attacks.
We run the attack as described in Section \ref{sec:practical} for different LWE parameter settings and report results in Table~\ref{tab:experiments_recovery}. After reduction, the running time depends almost entirely on the amount of enumeration that has to be done on the cruel bits.
Enumeration is done in ascending hamming weight on those bits.
We extend the secret to the cool bits via the greedy algorithm, every 40M secret candidates, or whenever completing the enumeration on one hamming weight.

All times reported in Table \ref{tab:experiments_recovery} include the compilation time during the first run over a secret, which is of the order of 10 seconds, and loading data, also of the order of 10 seconds. For some hamming weights, we report multiple timings to illustrate that the process is somewhat noisy, mostly depending on where in the enumeration the secret happens to be.
For the settings with higher hamming weights ($h=60$), we use 200k samples for the brute force attack and 1M for the greedy attack. For the lower hamming weights, we use 10k-30k samples for the brute force attack and 200k for the greedy part.  Recall that samples are generated via the pre-processing step explained in Section~\ref{sec:preprocess}, so the raw number of samples needed for each attack is always $4n$.  Also note that attacks are run on 1 V100 GPU in general,  using 20 GPUs in parallel only for the hardest cases.

\para{Existing attacks.} For completeness, we compare the performance of our attack to prior work. The hybrid dual meet-in-the-middle (MiTM) attack is most similar to ours~\cite{Cheon_hybrid_dual}, but options for comparison are limited, since no concrete performance results are provided in~\cite{Cheon_hybrid_dual}. Some concrete performance results are given for related attacks, but ~\cite{albrecht2017revisiting} only considers $n \le 110$, $\log q = 11$, while ~\cite{postlethwaite2021success} considers $n \le 100$, $\log q = 8$. Thus, we choose two routes for comparison. First, we compare against estimated cost of the hybrid dual MiTM estimate from the LWE estimator tool~\cite{LWEestimator}\footnote{https://github.com/malb/lattice-estimator, commit \texttt{00ec72ce}}, and then we compare against a recent concrete attack implementation for sparse secret LWE proposed in~\cite{rumpsession}. The latter has code available\footnote{https://github.com/lducas/leaky-LWE-Estimator/tree/human-LWE/human-LWE} and claims to be fast. It provides a MITM attack in Python for the $n=256$, $\log q=12$ setting and a modified uSVP/BDD attack for larger parameters.

\begin{table}[h]
\vspace{-0.6cm}
\centering
\caption{{\bf Claimed attacks for comparison from~\cite{LWEestimator} and~\cite{rumpsession}} \\
Estimates from~\cite{LWEestimator} do not represent actual secret recovery, only predictions. Estimates given in terms of ROP = estimated number of required operations for attack, 
Repeats = $\#$ of times attack must run to succeed with probability $0.99999$, Total Cost = total estimated ROP for this attack, omitting storage costs. For the code from~\cite{rumpsession}, `-' indicates no success after running for 3 weeks on our hardware.}
\vspace{0.1cm}
\label{tab:other_attacks}
    \setlength{\tabcolsep}{8pt}
\begin{tabular}{ccccc}
\toprule
\multirow{2}{*}{ 
\textbf{\begin{tabular}[c]{@{}c@{}}LWE setting\\ ($n,q,h$)\end{tabular}}} & \multicolumn{3}{c}{\textbf{Dual Hybrid MiTM~\cite{LWEestimator}}} & \textbf{Ducas et al}~\cite{rumpsession} \\ 
 & ROP & Repeats & Total Cost & Time (sec)  \\ \midrule
$(256,12,12)$ & $2^{52.7}$ & $2^{14.5}$ & $2^{67.2}$& 200 \\ 
$(512, 28,12)$ & $2^{48.6}$ & $2^{10.6}$ & $2^{59.2}$ &  - \\ 
$(512, 41, 60)$ & $2^{50.6}$ & $2^{8.6}$ & $2^{59.2}$ &  - \\ 
$(768, 35, 12)$ & $2^{49.1}$ & $2^{10.6}$ & $2^{59.7}$ & -  \\ 
$(1024, 50, 17)$ & $2^{50.9}$ &  $2^{11.1}$ & $2^{61.0}$ & -  \\ 
\bottomrule
\end{tabular}
\vspace{-0.2cm}
\end{table}

Table~\ref{tab:other_attacks} reports comparison results. For the estimator~\cite{LWEestimator}, we report the estimated number of operations and the number of attack repeats required to achieve a $1-10^{-5}$ probability of success (to match our $\alpha$, $\beta$ levels in \S\ref{sec:theory}). 
We note that the ROP metric reported by the estimator is a crude estimate for runtime, since it refers to the number of comparisons or operations necessary for the attack to run, which does not easily translate to time measurements. For example, if it means $(\mathbf{Z}/q\mathbf{Z})$  ring operations then one should multiply by the cost of a multiplication modulo $q$, naively $(\log q)^2$, or if it means ciphertext operations then one should multiply by the cost of polynomial multiplications modulo $q$, etc.

For the attack reported in~\cite{rumpsession}, we find that it does not succeed and does not recover secrets in the dimensions $ > 256$ reported in this paper and in~\cite{verde}. The attack ran for up to 3 weeks in all cases where dimension $> 256$ (on the same hardware as our attack). No secrets were recovered and in some cases the jobs crashed after running for days/weeks. This claimed attack and code was also presented in~\cite{RWCAlbrecht}, without verifying the code or the fact that it does not recover secrets in the dimensions ($>256$) we attack.


\section{2-power Cyclotomic Ring-LWE}
\label{sec:RLWE}

Our attack can also be applied in the 2-power cyclotomic Ring-LWE setting.
Ring-LWE (RLWE) is a special case of LWE where the LWE samples are represented more compactly as polynomials in a polynomial ring.
Not all instances of Ring-LWE are hard, as was shown in~\cite{ELOS}. But $2$-power cyclotomic rings are not vulnerable to the attack of~\cite{ELOS}, so those are the rings proposed for standardized use in Homomorphic Encryption~\cite{HES,bossuat2024security}. 

Consider the $2$-power cyclotomic ring defined by the polynomial $x^n+1$, where $n=2^k$: $R_q = \mathbb{Z}_q[x]/(x^n+1)$.
Then the RLWE samples $(a(x), b(x) = a(x) \cdot s(x) + e(x))$ can be described via a skew-circulant matrix, $\mathbf A_{circ}$.  
If $a(x) = a_0 + a_1x + ... + a_{n-1}x^{n-1}$, then using the embedding $\mathbf a=(a_0, a_1, \dots, a_{n-1})$:
\[
\mathbf A_{\text{circ}} =  \begin{bmatrix}
a_0 & -a_{n-1} & -a_{n-2} &\dots & -a_1 \\
a_1 & a_0 & -a_{n-1} & \ddots & -a_2 \\
\vdots & a_1 & a_0 & \ddots & -a_{n-1} \\
a_{n-1} & \ddots & \ddots & \ddots & a_0
\end{bmatrix}
\]
and $\mathbf b_{\text{circ}} = \mathbf A_{\text{circ}} \mathbf s + \mathbf e$ where $\mathbf s$ and $\mathbf e$ are the vector representations of $s(x)$ and $e(x)$ respectively.
\begin{figure}[t]
    \centering
    \includegraphics[width=0.5\linewidth]{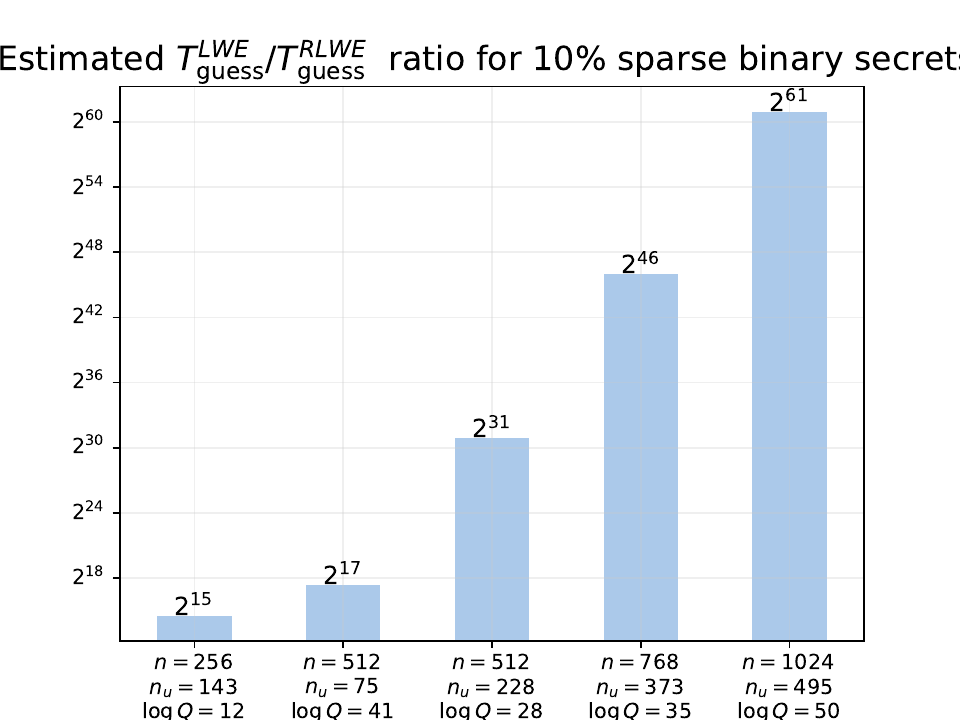}
    \vspace{-0.1cm}
    \caption{Exhaustive search cost ratio for LWE versus RLWE as defined in \cref{eq:tlwe} and \cref{eq:trlwe} assuming a fixed $10\%$ binary secret sparsity. The values of $n_u$ are obtained experimentally.}
    \label{fig:rlwe_tguess_ratio}
    \vspace{-0.4cm}
\end{figure}
When the matrix has this structure, we can rotate the ``cruel bits'' around (without redoing any lattice reduction) by constructing datasets composed of elements from certain indices in the matrix 
As Appendix Figure~\ref{fig:orig-shifted-hu} shows, shifting increases our chances of recovering the secret, since we can rotate the cruel region essentially for free.  Rotating around, we may find a rotation of the cruel region with only a few cruel (active) bits. We can then run our attack on this new, easier problem. This ability to shift the cruel region makes the RWLE problem clearly easier than generic LWE, with respect to our attack.

More formally, RLWE cruel bit shifting works as follows. After running reduction on the original RLWE polynomials $a(x)$, we have reduced samples,
 $\mathbf A \in \mathbb{Z}_q^{m\times n}$, where 
 $\mathbf A = \mathbf{RA}_{\text{circ}}$ for some  
 $\mathbf R\in \mathbb{Z}^{m\times n}$. 
 We can shift-negate the reduced samples $k$ times to get new samples  $(\mathbf A^{\leftarrow k}, \mathbf b^{\leftarrow k})$ (see \ref{sec:appx}). The attack for RLWE brute forces $n$ sliding windows $[k, k+n_u \mod{n})$ simultaneously (or any other number of windows by using a higher sliding step). Exhaustive search only requires checking up to $h^{(1)}_u = \min_{0 \leq i<n} h_{n_u, i}(\mathbf s)$ hamming weight where 
\begin{align}
    h_{n_u, i}(\mathbf s) = \sum_{j=i}^{i+n_u-1[n]} s_j
    \label{eq:hui}
\end{align}

 $h_{n_u, i}(\mathbf s)$ represents the Hamming weight of a specific segment of the secret vector $\mathbf s$, starting at position $i$ and of length $n_u$. When $i=0$, this notation aligns with our previous use of $h_u$ to denote the Hamming weight of the first $n_u$ elements of $\mathbf s$. Similarly, when $n_r=n-n_u$, $h_{n_r, n_u}(\mathbf s)$ aligns with our previous use of $h_r$ to denote the Hamming weight of the last $n_r=n-n_u$ elements of $\mathbf s$.

To evaluate the speed up of the RLWE attack, we estimate the cost of the brute force component while assuming equal cost of lattice reduction for LWE and RLWE. Let $c\cdot M(h_r(s), \alpha, \beta)$ represent the cost of checking a secret candidate using $M(h_r(s), \alpha, \beta)$ samples, where $c$ is a constant accounting for the computational cost of each verification. 
Let $T_{\text{guess}}^{\text{LWE}}$ be
the average time complexity of the exhaustive search for LWE, assuming we sweep through secret candidates in ascending Hamming weight order:
$$T_{\text{guess}}^{\text{LWE}} = \mathbb{E}_s\Biggl [c\cdot M(h_r(s), \alpha, \beta)\sum_{k=0}^{h_u(s) } \binom{n_u}{k}\Biggr]$$

\noindent Using the same number of samples $M=M(h_r(s), \alpha, \beta)$ for each secret candidate:
\begin{equation}
    \label{eq:tlwe}
    T_{\text{guess}}^{\text{LWE}} = c\cdot M\cdot \sum_{k=0}^{h} \binom{n_u}{k}\mathbb{P}(h_u(s) \geq k)
\end{equation}

\noindent Contrast this with the RLWE case, where we have $\mathbb{P}(h^{(1)}_u \leq \frac{hn_u}{n}) = 1$ so that: 
\begin{equation}
    \label{eq:trlwe}
    T_{\text{guess}}^{\text{RLWE}}  = n\cdot c\cdot M\cdot \sum_{k=0}^{\lfloor \frac{hn_u}{n}\rfloor} \binom{n_u}{k}\mathbb{P}(h^{(1)}_u \geq k)
\end{equation}

In \cref{fig:rlwe_tguess_ratio}, we estimate the ratio of these costs. The distribution of $h_u(s)$ with fixed total hamming weight $h$ follows a hyper-geometric distribution when the secret bits are from a Bernoulli distribution, while the $h^{(1)}_u$ distribution is computed empirically.

\vspace{-0.3cm}
\begin{table}[h]
\vspace{-0.4cm}
    \caption{{\bf RLWE/LWE Attack times} for $n = 256, h = 12, \log q = 12, n_u = 145$, run on 10 randomly sampled secrets. Actual time is the empirical time of the attack. Estimated time is computed via the formula: $T_{\text{guess}}^{\text{LWE}} = b + a\sum_{k=0}^{h_u(s)} \binom{n_u}{k}$ and $T_{\text{guess}}^{\text{RLWE}} = n(b + a\sum_{k=0}^{h^{(1)}_u} \binom{n_u}{k})$ for RLWE. 
    $h_u(s) = \# $ cruel bits of the secret,
     and $h^{(1)}_u$ is the minimum hamming weight over all windows of length $n_u$. We use $56K$ samples for cruel bit recovery and $200K$ for \cref{alg:greedy_cool}. }
    \vspace{0.1cm}
    \centering
    \small
    \setlength{\tabcolsep}{5pt}
    \begin{tabular}{ccccccc}
\toprule
 & \multicolumn{3}{c}{LWE}  & \multicolumn{3}{c}{RLWE}  \\
\cmidrule{2-7}
 Secret & Estimated & Actual & &  Estimated &Actual &  \\
 & time (sec) & time (sec) & $h_u(s)$ &    time & time&$h^{(1)}_u(s)$\\
\midrule
1 & $2.33\times 10^1$ & $2.21\times 10^1$ & 3 & $5.86\times 10^3$ & $4.83\times 10^3$ & 2 \\
2 & $4.60\times 10^2$ & $5.24\times 10^2$ & 5 & $1.18\times 10^5$ & $1.73\times 10^5$ & 5 \\
3 & $2.09\times 10^5$ & -- & 7 & $1.18\times 10^5$ & $1.48\times 10^5$ & 5 \\
4 & $2.09\times 10^5$ & -- & 7 & $9.77\times 10^3$ & $1.20\times 10^4$ & 4 \\
5 & $2.09\times 10^5$ & -- & 7 & $9.77\times 10^3$ & $1.15\times 10^4$ & 4 \\
6 & $3.66\times 10^6$ & -- & 8 & $9.77\times 10^3$ & $1.26\times 10^4$ & 4 \\
7 & $3.66\times 10^6$ & -- & 8 & $9.77\times 10^3$ & $1.19\times 10^4$ & 4 \\
8 & $5.65\times 10^7$ & -- & 9 & $5.97\times 10^3$ & $4.12\times 10^3$ & 3 \\
9 & $5.65\times 10^7$ & -- & 9 & $1.18\times 10^5$ & $1.33\times 10^5$ & 5 \\
10 & $7.81\times 10^8$ & -- & 10 & $5.86\times 10^3$ & $4.74\times 10^3$ & 2 \\
\bottomrule
\end{tabular}
\label{tab:n256}
\vspace{-0.3cm}
\end{table}

We conduct several experiments comparing LWE with RLWE, using identical secrets. The results of these experiments, for parameters $n=256$, $n=512$ and $n=1024$, are presented in Tables \ref{tab:n256}, \ref{tab:n512}, and \ref{tab:n1024}. Each table row corresponds to a distinct secret $s$, and the columns detail the actual/estimated time costs for the LWE/RLWE attacks, along with the number of cruel bits in the secret. In the tables, Actual Time is the empirically measured duration of the attack, while the Estimated Time is estimated using an approximation $T_{\text{guess}}^{\text{LWE}} = b + a\sum_{k=0}^{h_u(s)} \binom{n_u}{k}$. This estimate provides an idea of the time required to attack a secret when $h_u$ is large. 

For RLWE attacks, time is multiplied by $n$ since $n$ brute force attacks are run, although they can be run simultaneously. The final row of Table~\ref{tab:n256} provides the average values across all secrets. As can be seen, the RLWE attack is significantly faster on average, with the ratio of average times being $\sim 10^3$ for $n=256$ and $5\times 10^3$ for $n=512$, but converge to the theoretical values in \cref{fig:rlwe_tguess_ratio} when estimated on more secret samples. The RLWE advantage can also be seen in the values of $h^{(1)}_u(s)$ which are much smaller than $h_u(s)$. 

\vspace{-0.5cm}
\begin{table}[h]
\vspace{-0.3cm}
    \caption{Attack times vs. estimated costs for $n=512, n_u=75, h=60, \log q = 41$ (in seconds) for $10$ random secrets sorted by $\#$  cruel bits in the secret. We use $40K$ samples for cruel bit recovery and $1.5M$ for \cref{alg:greedy_cool} }
    \vspace{0.1cm}
    \centering
    \small
    \setlength{\tabcolsep}{5pt}
    \begin{tabular}{ccccccc}
    \toprule
     & \multicolumn{3}{c}{LWE}  & \multicolumn{3}{c}{RLWE}  \\
    \cmidrule{2-7}
     Secret & Estimated & Actual & &  Estimated &Actual &  \\
     & time (sec) & time (sec) & $h_u(s)$ &    time & time&$h^{(1)}_u(s)$\\
    \midrule
    1 & $6.88\times 10^2$ & $5.95\times 10^2$ & 6 & $1.82\times 10^5$ & $7.74\times 10^4$ & 4 \\
    2 & $3.98\times 10^3$ & $3.93\times 10^3$ & 7 & $1.82\times 10^5$ & $7.68\times 10^4$ & 4 \\
    3 & $3.45\times 10^4$ & -- & 8 & $3.52\times 10^5$ & $1.08\times 10^5$ & 6 \\
    4 & $2.82\times 10^5$ & -- & 9 & $1.95\times 10^5$ & $1.15\times 10^6$ & 5 \\
    5 & $2.06\times 10^6$ & -- & 10 & $1.82\times 10^5$ & $4.64\times 10^4$ & 4 \\
    6 & $2.06\times 10^6$ & -- & 10 & $1.81\times 10^5$ & -- & 2 \\
    7 & $1.36\times 10^7$ & -- & 11 & $1.82\times 10^5$ & -- & 4 \\
    8 & $8.06\times 10^7$ & -- & 12 & $1.81\times 10^5$ & $4.43\times 10^4$ & 3 \\
    9 & $4.36\times 10^8$ & -- & 13 & $1.82\times 10^5$ & -- & 4 \\
    10 & $9.88\times 10^9$ & -- & 15 & $1.81\times 10^5$ & $4.67\times 10^4$ & 2 \\
    \bottomrule
    \end{tabular}
    \label{tab:n512}
    \vspace{-0.3cm}
\end{table}
\vspace{-0.3cm}

\vspace{-0.5cm}
\begin{table}[h]
\vspace{-0.3cm}
    \caption{Attack times vs. estimated times for $n=1024, h=20, \log q = 50, n_u=495$ (in seconds) for $10$ random secrets sorted by $\#$  cruel bits in the secret. We use $10K$ samples for cruel bit recovery (scaled to 4 GPUs) and $100K$ for \cref{alg:greedy_cool} }
    \vspace{0.1cm}
    \centering
    \small
    \setlength{\tabcolsep}{5pt}
    \begin{tabular}{ccccccc}
    \toprule
     & \multicolumn{3}{c}{LWE}  & \multicolumn{3}{c}{RLWE}  \\
    \cmidrule{2-7}
     Secret & Estimated & Actual & &  Estimated &Actual &  \\
     & time (sec) & time (sec) & $h_u(s)$ &    time & time&$h^{(1)}_u(s)$\\
    \midrule
    1 & $6.34\times 10^9$ & -- & 8 & $1.51\times 10^9$ & -- & 6 \\
2 & $6.34\times 10^9$ & -- & 8 & $2.01\times 10^7$ & $2.01\times 10^7$ & 5 \\
3 & $3.45\times 10^{11}$ & -- & 9 & $1.71\times 10^6$ & $4.78\times 10^5$ & 3 \\
4 & $3.45\times 10^{11}$ & -- & 9 & $1.51\times 10^9$ & -- & 6 \\
5 & $3.45\times 10^{11}$ & -- & 9 & $1.06\times 10^{11}$ & -- & 7 \\
6 & $1.68\times 10^{13}$ & -- & 10 & $1.06\times 10^{11}$ & -- & 7 \\
7 & $1.68\times 10^{13}$ & -- & 10 & $1.51\times 10^9$ & -- & 6 \\
8 & $1.68\times 10^{13}$ & -- & 10 & $1.06\times 10^{11}$ & -- & 7 \\
9 & $3.02\times 10^{16}$ & -- & 12 & $1.51\times 10^9$ & -- & 6 \\
10 & $1.13\times 10^{18}$ & -- & 13 & $1.90\times 10^6$ & $3.14\times 10^6$ & 4 \\
    \bottomrule
    \end{tabular}
    \label{tab:n1024}
    \vspace{-0.1cm}
\end{table}

\section{Discussion and Future Work}

We have presented an attack on Learning With Errors in the setting of sparse binary secrets. Our key insight is the reduction of the LWE problem to one of smaller dimension via lattice reduction and sufficient subsampling.
We discuss the attack empirically and theoretically and highlight actual successful attacks for dimension up to $n=1024$. We leave several important tasks as future work.

\para{Improving upon combinatorial scaling in reduced problem.} We use brute force to solve the smaller LWE problem (e.g. in the cruel region).
This is the crudest way to solve this problem, albeit perfectly parallelizable. The guessing parts of hybrid attack strategies of related work could speed up enumeration. One related way in which we already exploit this is in the 2-cyclotomic RLWE setting.
But of course, the rotation (or more generally, permutation) of the cruel bit region can of course be done on the LWE settings as well. We did not elaborate on that (hybrid) approach because it requires re-reduction of lattices.

\para{Balancing reduction and secret recovery costs.} Currently, our attack runs lattice reduction until it stalls for a given $n$/$q$ setting, then runs secret recovery for the highest $h$ that can be achieved in reasonable time. In the future, our attack should balance the reduction and enumeration costs given the setting of concern, also depending on how hybrid attacks will be employed.

\para{Recovering the cool bits.} As Section~\ref{sec:cool_bits} demonstrates, cool bit recovery is somewhat sample-inefficient, because \cref{alg:greedy_cool} must distinguish between $F_q(k\sigma_r^2+\sigma_e^2)$ and $F_q((k+1)\sigma_r^2+\sigma_e^2)$. This difference tends to be small when $k$ is large, and $\sigma_e$ is large, compared to $\sigma_r$, necessitating more samples. This suggests an additional consideration for the lattice reduction phase: balancing the amount of reduction achieved and the noise added in the process. When reduction factor $\rho$ is smaller, $\sigma_r$ is smaller, resulting in fewer cruel bits and easier recovery. On the other hand, a smaller $\rho$ means a larger $\sigma_e$, necessitating more data during the attack. We leave the quantitative analysis of this relationship as future work.


\appendix
\section{Appendix}

\subsection{Rotating Reduced Short 2-cyclotomic RLWE Vectors}
\label{sec:appx}
Consider the $R_q$-endomorphism $\phi$ defined by $\phi: a \mapsto xa$. We establish its counterpart in the canonical embedding as:
\[
\Phi(A) = AX, \quad \forall A \in \mathbb{Z}^{m\times n},
\]
where $X$ is defined as $X = \text{Circ}(x_{\text{vec}})^\top$, with Circ denoting the skew-circulant and $x$ being the degree 1 polynomial whose vector representation is given by $x_{\text{vec}} = (0, 1, 0, \dots, 0) \in \mathbb{Z}^n$. We further introduce the notation $\Phi_k(A)$ to denote the $k$-fold composition of $\Phi$, i.e., $\Phi^k(A) = \Phi \circ \Phi \circ \dots \circ \Phi$. If $A_{\text{circ}} = \text{Circ}(a)$, then $A_{\text{circ}} $ and $X$ commute: $XA_{\text{circ}} = A_{\text{circ}} X$. This is due to the associativity in the ring $R_q$. This allows us to have new samples $(A^{\rightarrow k}, b^{\rightarrow k})$ by shift-negating the reduced samples $k$ times:

\begin{minipage}{0.44\textwidth}
\[
    A^{\rightarrow k} := AX^k = \Phi_k(A) \quad \text{ and}
\]
\end{minipage}
\hfill
\begin{minipage}{0.44\textwidth}
\begin{align*}
    A^{\rightarrow k}s &= RA_{\text{circ}}X^ks \\
    &= RX^kA_{\text{circ}}s \\
    &= RX^k(A_{\text{circ}}s + e) -  RX^ke\\
    &= \Phi_k(R)b_{\text{circ}} - \Phi_k(R)(e)
\end{align*}
\end{minipage}

\noindent So by defining 
    $b^{\rightarrow k} := R^{\rightarrow k}b_{\text{circ}} 
$, 
$(A^{\rightarrow k}, b^{\rightarrow k})$ are LWE samples with secret $s$.

\begin{figure*}[h]
    \centering
    \includegraphics[width=0.48\textwidth]{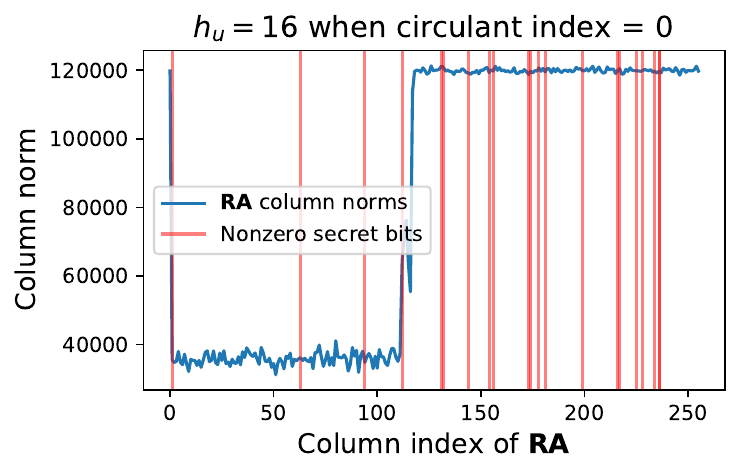}\hfill
    \includegraphics[width=0.49\textwidth]{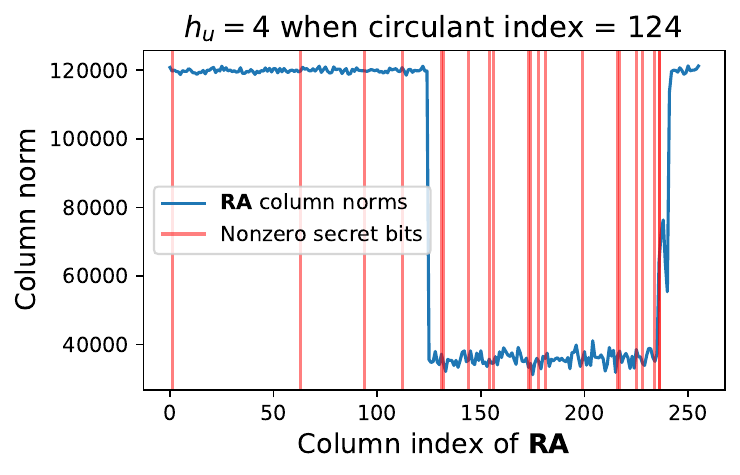}
    \vspace{-0.2cm}
    \caption{Effect of shifting the RLWE circulant matrix on number of $h_u$ secret bits, for $n=256$, $\log_2=12$, $n_u=143$, $h=20$.}
    \label{fig:orig-shifted-hu}
    \vspace{-0.3cm}
\end{figure*}

We illustrate empirically how sub-selecting samples from different indices of the circulant matrix affects $h_u$ in Figure~\ref{fig:orig-shifted-hu}. When we sub-sample elements when $k=0$ (e.g. no shifting), $h_u=16$ for a $n=256$ $\log q=12$, $h=16$ problem. However, sub-sampling elements at $k=124$ yields $h_u=4$, a much easier secret.

\vspace{-0.3cm}

\subsection{Minimum Sample Requirement for Hypothesis Testing}
\label{sec:m_formula_dx}

We examine two centered discrete distributions, namely $\mathcal{U}_q := \textit{Uniform}(\mathbb{Z}_q)$ and $\mathcal{G}_q:= \textit{Discrete Gaussian}(0, \sigma_{\mathcal{G}}^2) \mod q$ where $\sigma_{\mathcal{G}}^2 = h_r\sigma_r^2 + \sigma_e^2$ and $\sigma_{\mathcal{G}_q}^2 =F_q(\sigma_{\mathcal{G}}^2)$.

The variance of $x^2$ when $x$ is uniform is $\sigma^2_{\mathcal{U}_q^2} = \sigma^2_{\mathcal{U}_q}\frac{q^2-4}{15}$. For the discrete Gaussian $\mod q$,  the variance of $x^2$: $\sigma^2_{\mathcal{G}_q^2} 
 = \mathbb{E}_{\mathcal{G}_q}(x^4)-\mathbb{E}_{\mathcal{G}_q}(x^2)^2$ is approximated. Let $x$ be a sample drawn from either of the two aforementioned distributions. We also consider $M$ samples $x_i$ drawn from the same distribution as $x$. To differentiate between the two distributions, we conduct a statistical test with the following null and alternative hypotheses:
\begin{align*}
    \textbf{H0: } \sigma^2(x) = \sigma^2_{\mathcal{U}_q} \quad
\textbf{H1: } \sigma^2(x) = \sigma_{\mathcal{G}_q}^2
\end{align*}

We consider the unbiased variance estimator for a known 0 mean: $\hat{\sigma}^2_M = \frac{1}{M}\sum_{i=0}^{M-1} x_i^2$. The threshold $\iota(\alpha, M)$ is set so that the type-1 error: $\alpha = \mathbb{P}_{\mathcal{U}_q}(\hat{\sigma}^2_M < \iota(\alpha, M)) = \mathbb{P}(\hat{\sigma}^2_M < \iota(\alpha, M)| {x\sim \mathcal{U}_q})$ is given. Using Central Limit Theorem and solving for $\iota(M, \alpha)$
\begin{align*}
    \alpha = \mathbb{P}(\hat{\sigma}^2_M < \iota| {x\sim \mathcal{U}_q}) &= \mathbb{P}(\sqrt{\frac{M}{\sigma^2_{\mathcal{U}_q^2}}}(\hat{\sigma}^2_M-\sigma^2_{\mathcal{U}_q})  < \sqrt{\frac{M}{\sigma^2_{\mathcal{U}_q^2}}}(\iota-\sigma^2_{\mathcal{U}_q})| {x\sim \mathcal{U}_q}) \\
    &\approx \mathbb{P}(\mathcal{G}(0,1) < \sqrt{\frac{M}{\sigma^2_{\mathcal{U}_q^2}}}(\iota-\sigma^2_{\mathcal{U}_q})) =G(\sqrt{\frac{M}{\sigma^2_{\mathcal{U}_q^2}}}(\iota-\sigma^2_{\mathcal{U}_q}))
\end{align*}

Which results in the equation: 
\begin{align}
    \label{equation:iota}
     \iota(M, \alpha) &=   \sigma^2_{\mathcal{U}_q} + G^{-1}(\alpha)\frac{\sigma_{\mathcal{U}^2_q}}{\sqrt{M}}
\end{align}
Doing the same for the type-2 error given $\iota(M, \alpha)$ and solving for $M$:

\begin{align*}
    \beta &= \mathbb{P}(\hat{\sigma}^2_M \geq \iota(M, \alpha)| {x\sim \mathcal{G}_q})
    \approx  G\Big (\sqrt{\frac{M}{\sigma^2_{\mathcal{G}_q^2}}}(\sigma^2_{\mathcal{G}_q}-\iota(M, \alpha))\Big)
\end{align*}
By substituting $\iota(M, \alpha)$ by its expression in \cref{equation:iota}, we have:

\begin{align*}
    M =  \Biggl [\frac{ G^{-1}(\alpha) \sigma_{\mathcal{U}_q^2} + G^{-1}(\beta)\sigma_{\mathcal{G}_q^2}}{(\sigma^2_{\mathcal{G}_q} - \sigma^2_{\mathcal{U}_q})} \Biggr ]^2 
\end{align*}

\subsection{Our Observation and the q-ary Lattice Z-shape}
\label{subsec:zshape}
\begin{figure}[h!]
    \centering
    \vspace{-0.5cm}
    \includegraphics[width=0.9\textwidth]{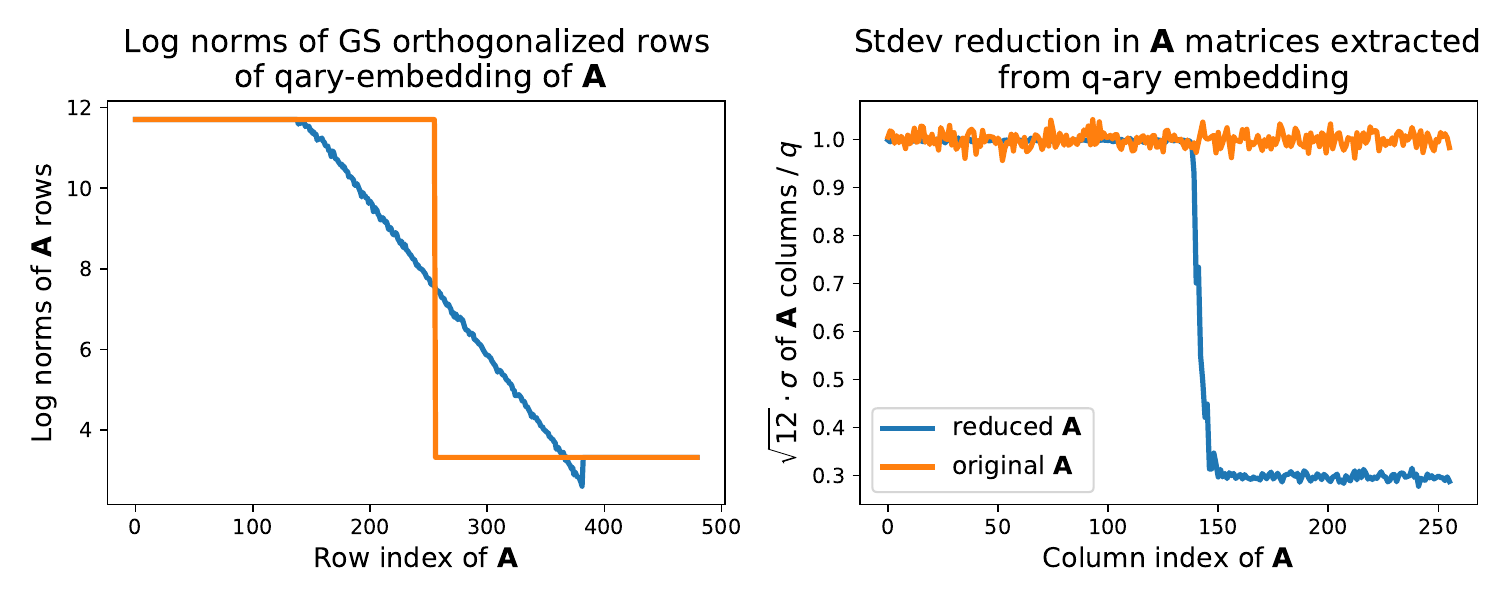}
    \caption{
    } 
    \label{fig:profile_comparison}
    \vspace{-0.5cm}
\end{figure}

Prior work has observed a "z-shape" in q-ary lattices in the context of the hybrid attack~\cite{howgrave2007hybrid,albrecht2021lattice}. This classic z-shape is exhibited by the Gram-Schmidt orthogonalized rows of a q-ary lattice (see left plot of Figure~\ref{fig:profile_comparison}, orange before reduction, blue after). This is distinct from the behavior we observe in the columns of $\mathbf{A}$ after reduction (see right plot of~\cref{fig:profile_comparison} in blue).

\bibliographystyle{splncs04}
\bibliography{references}

\end{document}